\input harvmac
%\draftmode

\def\R{\relax{\rm I\kern-.18em R}}
\font\cmss=cmss10 \font\cmsss=cmss10 at 7pt
\def\Z{\relax\ifmmode\mathchoice
{\hbox{\cmss Z\kern-.4em Z}}{\hbox{\cmss Z\kern-.4em Z}}
{\lower.9pt\hbox{\cmsss Z\kern-.4em Z}}
{\lower1.2pt\hbox{\cmsss Z\kern-.4em Z}}\else{\cmss Z\kern-.4em Z}\fi}\
\def\np{{\it Nucl. Phys.}}
\def\pl{{\it Phys. Lett.} }

\def\prl{{\it Phys. Rev. Lett. }}
\def\p{\partial}

\def\CC{{\cal C}}

\def\l{\lambda}

\def\p{\partial}

\def\tr{{\rm tr}}
 %references
 \lref\ksw{V. Kazakov, M. Staudacher and T. Wynter, LPTENS-95/9;  
LPTENS-95/56}  
\lref\dkaz{F. David, \np B 257 (1985) 45;
V. Kazakov, \pl B 150 (1985) 282.}
\lref\bk{
 V.Kazakov, \pl  119 A (1986) 140;
D. Bulatov and  V.Kazakov, \pl 186 B (1987) 379.}
\lref\onon{I. Kostov, {\it Mod. Phys. Lett.} A4 (1989) 217;
 M. Gaudin and I. Kostov, \pl B 220 (1989) 200; I. Kostov and M. 
Staudacher, \np B  384 (1992) 459; B. Eynard and J. Zinn-Justin, \np B 386 
(1992) 55.}  
 \lref\kjm{
 V. Kazakov, \np B (Proc. Supp.) 4 (1988) 93;
J.-M. Daul, Q-states Potts model on a random planar lattice,
 hep-th/9502014.}
  \lref\kmul{
 V. Kazakov, {\it Mod. Phys. Lett.} A4 (1989) 2125.}
\lref\Iade{
  I. Kostov, \np B 326 (1989)583;  \np B 376 (1992) 539.}
 \lref\adem{I. Kostov,  \pl  297 (1992) 74.}
 \lref\higkos{
S. Higuchi and I. Kostov,   \pl B 357 (1995) 62.}
 \lref\berchar{
 B. Eynard and C. Kristjansen, preprints SPhT/95-068, 
SPhT/95/133.  }
\lref\phil{ P. Di Francesco and D. Kutasov, \np  B 342  (1990) 589}
 \lref\sixauthors{
M. Douglas and S. Shenker, \np B 335 (1990) 635;
E. Br\'ezin and V. Kazakov, \pl B 236 (1990);
D. Gross and A. Migdal, \prl 64 (1990) 127 ; \np B 340 (1990) 333.}
 \lref\mike{
M. Douglas, \pl  B 238 (1990) 176;
M. Fukuma, H. Kawai and R. Nakayama, { \it Int. J. Mod.
Phys. }  A 6 (1991) 1385;
R. Dijkgraaf, H. Verlinde and E. Verlinde, \np B 348 (1991) 435.}
 \lref\Itrieste{
 I. Kostov,  Solvable Statistical Models on Random Lattices,
Proceedings of the "Conference on recent developments in statistical
mechanics and quantum field theory"  (10 - 12 April 1995), Trieste, Italy.}
 \lref\alfak{
J. Alfaro and I. Kostov, to be published.} 
\lref\KMMM{
S. Kharchev, A. Marshakov, A. Mironov and A. Morozov, 
\np B 397 (1993) 339 and references therein.}
\lref\hirota{
R. Hirota, \prl  27 (1971) 1192.}
 \lref\jimiwa{
 M. Jimbo and T. Miwa, "Solitons and infinite dimensional Lie algebras",
{\it RIMS } Vol. 19 (1983) 943-1001.}
 \lref\mehta{
M. L. Mehta, $Random Matrices$  (second edition), Academic Press, New 
York, 1990.}
 \lref\izii{
C. Itzykson and J.-B. Zuber, J. Math. Phys. 21 (1980) 411.}
 \lref\zamsal{ 
P. Fendley and H. Saleur, \np B 388 (1992) 609,
 Al. B. Zamolodchikov, \np B 432 [FS] (1994) 427.}
\lref\trw{ 
D. Bernard and A. LeClair, \np B426 (1994) 534;
 C. Tracy and H. Widom, Fredholm determinants and
the mKdV/sinh-Gordon hierarchies, solv-int/9506006, S. Kakei, Toda Lattice 
Hierarchy and Zamolodchikov conjecture, solv-int/9510006.}
\lref\DS{Drinfeld and V. Sokolov, {\it. Itogi Nauki i Techniki} 24 (1984) 
81,
{\it Docl. Akad. Nauk. SSSR} 258 (1981) 1.}
\lref\DFK{P. Di Francesco and D. Kutasov, \np B 342 (1990) 589.}
 \lref\KW{V. Kac and M. Wakimoto, Exceptional hierarchies of soliton 
equations, {\it
Proceedings of Symposia in Pure Mathematics}, 49 (1989) 191}
  \Title{}
{\vbox{\centerline
{   Bilinear Functional Equations   }
\vskip2pt
\centerline{ in  2D   Quantum Gravity   
    }}}

\vskip6pt

\centerline{Ivan K. Kostov \footnote{$ ^\ast $}{on leave of absence
from the Institute for Nuclear Research and Nuclear Energy,
Boulevard Tsarigradsko shosse 72, BG-1784 Sofia,
 Bulgaria}\footnote{$ ^\dagger$}{{\tt kostov@amoco.saclay.cea.fr}}}

\centerline{{\it C.E.A. - Saclay, Service de Physique Th\'eorique}}
 \centerline{{\it 
CE-Saclay, F-91191 Gif-Sur-Yvette, France}}

\vskip .3in
\baselineskip10pt{ 
The microscopic theories of quantum gravity related to integrable lattice models
can be constructed as special deformations of pure gravity. Each such deformation
is defined by a second order differential operator acting on the coupling constants.
As a consequence, the theories with matter fields satisfy a set of constraints 
inherited from the integrable structure of pure gravity. In particular, a set of 
bilinear functional equations for each theory with matter fields follows from the
Hirota equations defining the KP (KdV) structure of pure gravity.    
      }

\vskip 1cm
\centerline{    Talk 
delivered at the Workshop on "New Trends in Quantum Field Theory"}
\centerline{  (28 August - 1 September 1995),   Razlog, Bulgaria.
    }
 
\bigskip
\rightline{SPhT-96/011}
%\draft
\Date{February 1996 }

\baselineskip=16pt plus 2pt minus 2pt
   
%%%%%%%%%%%%%%%%%%%%%%%%%%%%%%%%%%%%%%%%%%%%%%%%%%%%%%%%%%%%%%%%%%%%%%%%%%%%

\newsec{Introduction}

The simplest and perhaps the only possible microscopic realization of the 
two-dimensional quantum gravity is 
provided by the  ensemble of planar graphs \dkaz . 
In its turn,   the ensemble of 
planar graphs  is generated by the (divergent)  quasiclassical expansion 
of  $N\times N$ matrix integrals,  with the parameter  $1/N$ 
playing the role of  topological coupling constant. 
The formulation of 2D quantum gravity in terms of  random matrix variables 
opened the possibility to apply  powerful 
nonperturbative techniques  as  the method of orthogonal polynomials \izii 
\
and led to the discovery of unexpected integrable structures 
associated with its  different scaling regimes.  Originally, a structure 
related 
to the KdV  hierarchy of soliton equations  was found  in the continuum 
limit 
of pure gravity \sixauthors .  The KdV hierarchy  can be obtained as 
  a 2-reduction of the KP hierarchy, which defines the integrable  
structure
 of the microscopic theory.   Consequently, the partition functions of the
  $A$ series of models of 
matter  coupled to 2D gravity  were identified as $\tau$-functions of 
higher reductions of the KP integrable hierarchy \mike .  Similar 
statement concerning the  $D$ series was made in  \phil .

The description of interacting gravity and matter based on the   higher
 reductions of the  KP hierarchy 
is well suited  for studying  the spectroscopy and the correlation
 functions of local  scaling operators.   On the other hand, it   becomes 
less 
  efficient for evaluating  macroscopic loop correlators,   topology 
changing 
  interactions  and    other infrared phenomena.
Indeed, at distances much larger than the correlation length of the matter 
fields, the relevant integrable structure is this of pure gravity.
 Therefore it make sense to look for   a complementary  description  in 
which 
 the matter is considered as a perturbation 
 of pure gravity. The simplest realization of this idea is given by 
 the Kazakov's multicritical points of the one-matrix model \kmul , which 
 describe a series of nonunitary matter fields  coupled to gravity.
  
 A systematic  construction of the matter fields  as deformations
  of pure gravity  will be the  subject of this talk,
  which is partially based on the work \alfak .    Our main task 
  will be to  exhibit in any theory with matter fields the integrable
  structure inherited from pure gravity. Namely, we are going to derive 
   bilinear functional equations that can be viewed as a  
   deformation of the  Hirota equations for the partition function of 
   pure gravity. 
   We are going to  consider  in details  only  the microscopic realization
    of the theory, but  our   construction survives without substantial 
changes  in the continuum limit.

   We start with  the derivation of the Virasoro 
constraints  and the Hirota bilinear equations for the matrix  integral 
 generating
the ensemble of planar graphs. 
 The Virasoro constraints  follow from  the loop equations, which relate 
boson-like quantities (traces), while the   Hirota equations follow from 
orthogonality  relations involving 
 fermion-like quantities (determinants). 
Then we  show how both types of equations generalize for the 
matrix integrals describing theories of gravity with matter fields.
For this purpose we will exploit the microscopic construction of    $2D$ 
quantum 
gravity    given by the  random-lattice versions of the $sl(2)$-related  
statistical models: the Ising  \bk \ and the     $O({\rm n})$  \onon\ 
models, 
the   SOS and RSOS models and their  
    $ADE$ and $\hat A \hat D\hat E$ generalizations \Iade -\adem 
    \foot{The $q$-state Potts \kjm\ needs a more elaborate 
  construction and will not be considered here.}.  
   Each of these  models can be reformulated as theory  of one or several 
     random matrices with interaction  of the   form 
     $\tr\ln (M_a\otimes 1+1\otimes M_{b})$.  
  Such an  interaction   can be introduced by means of  a   differential 
operator
   of  second order ${\bf H}$ acting on the coupling constants;   the  
    partition function of the model is obtained by acting with the operator
      $e^ {\bf H}$ on  the  partition function of one or several decoupled 
       one-matrix models.   The operator $e^{\bf H}$ defines a canonical 
       transformation of the symplectic structure associated with the 
       space of  coupling constants.

 The Virasoro constraints $L_n^a=0$  for each of the one-matrix integrals  
 transform  to linear differential constraints $e^ {\bf H} L_n ^ae^ {-\bf 
H}=0$
  for the   interacting theory. These constraints are equivalent to  the 
loop 
  equations, which have been 
already derived by other means. The new point  is that the Hirota bilinear 
equations 
that hold for each of the one-matrix integrals induce, by simply replacing 
the vertex operators as  $ {\bf V}_{\pm}(z)^a \to
  e^ {\bf H}  {\bf V}_{\pm}^a(z) e^ {-\bf H}$,  bilinear functional 
equations for the interacting theory.    In 
this way the integrable structure
 associated with pure gravity is  deformed but not destroyed  by the
   matter fields.  In particular, the deformed Virasoro and vertex 
operators satisfy the same algebra as the bare ones.

 In the continuum limit,  the loop equations and the bilinear 
equations are obtained as deformations of the Virasoro constraints and 
the Hirota equations in the KdV hierarchy. Their  perturbative solution   
is given by the loop-space Feynman   
rules obtained   in \higkos .

   \newsec{The one-matrix model}

 The partition function of the ensemble of all two-dimensional 
  random  lattices  is  given  by the hermitian $N\times N$  matrix 
integral
 \eqn\hemm{
Z_N[t]\sim  \int dM \exp \Big( {\rm tr} \sum_{n=0}^{\infty} t_n 
M^n\Big) .}
The integrand depends on the matrix variable $M$ only through its  
eigenvalues 
$\lambda_i,\ 
i=1,...,N$.
Retaining, therefore, only the 
radial part of the integration measure $  dM \sim \prod _{i=1}^N d\l_i 
\prod_{i\ne j} (\l_i-\l_j)$ we can write the integral   
\hemm\  as the  partition 
 function for  a system of $N$ Coulomb particles in a common potential  
 \eqn\partf{
Z_{N}[t]= \int   \prod _{i=1}^N    d\lambda_i \ \exp 
\Big( \sum_{n=0}^{\infty} t_n \lambda^n_i\Big)  \ \prod_{i< j} (\l_i-\l_j) 
^2 .}
Let us recall that the last factor is the square of   the Vandermonde 
determinant
 \eqn\vander{\Delta_N(\lambda_1,...,\lambda_N)=\prod_{1\leq i<j \leq 
N} (\lambda_i - \lambda_j) = \det_{ij} [(\lambda_i)^{j-1}].}
   In the following we will denote by 
 $\langle ... \rangle_{N, t}$ the mean value defined by the  partition 
function  \partf .  
 For the time being we restrict the integration in $\lambda$'s to
    a  finite 
interval $[\lambda_L, \lambda_R]$  on the real axis, so that the measure
\eqn\meas{d\mu_t(\lambda)=d\lambda \exp 
\Big(\sum_{n=0}^{\infty}t_n\lambda^n\Big)}
is integrable for any choice of the coupling constants $t_n$.
  The limit $\l_{L}\to -\infty , \l_R\to\infty$  exists if the exponential 
$e^{\sum_n t_n \lambda^n}$
vanishes at $\pm\infty$.

\subsec{Loop equations and Virasoro constraints}

 The loop equations determine  the   $1/N$ expansion of the 
integral \partf\ and  represent an infinite set of identities satisfied by 
the 
 correlation functions of the collective  loop  variable 
  \eqn\www{W(z)=\sum_{i=1}^N {1\over z-\l_i} = \tr\Big( {1\over z-M}\Big).}
The    derivation goes as  follows.   From the translational invariance of
 the integration measure $d\lambda_i$ 
we find, neglecting the boundary terms at $\l = \l_{L,R}$,
\eqn\dseon{ 
\Big\langle \sum_{i=1}^N  \Big({\p\over\p\lambda_i}+
2\sum_{j(\ne i)} {1\over \lambda_i-\lambda_j}  
+\sum_{n\ge 0}nt_n\lambda_i^{n-1} \Big){1\over z-\lambda_i} 
\Big\rangle_{N,t}  =0. }
  Using the identity
\eqn\sdaa{
\sum_i {1\over (z-\lambda_i)^2}+2\sum _{i\ne j} {1\over 
z-\lambda_i}{1\over \lambda_i-\lambda_j}=
\sum_{i,j} {1\over z-\lambda_i} {1\over z-\lambda_j}}
we find
\eqn\dsa{ 
\Big\langle W^2(z)+  
\sum_{i=1}^N {1\over z-\lambda_i} \sum_{n\ge 0}nt_n\lambda_i^{n-1}  
\Big\rangle_{N,t} =0.}
 The sum in the second term can be viewed as the result of  a 
contour integration 
   along a   contour $\CC$   enclosing the interval 
$[\l_L,\l_R]$   but leaving outside   the point $z$:  
\eqn\eqct{   \Big\langle  W(z)^2  +  \oint_{\CC}
{dz'\over 2\pi i} { W(z') \over z-z'}  \sum_n
n(z')^{n-1}t_n\Big\rangle_{N,t} =0.}
 Introducing the collective field
 \eqn\colf{\Phi(z)={1\over 2}  \sum_{n\ge 0}
t_n z^ n -\tr \log \Big({1\over z-M}\Big) }
we write the last equation as 
\eqn\eqct{  \oint_{\CC}
{dz'\over 2\pi i} {1\over z-z'} \Big\langle \big(\p \Phi(z')\big)^2  
\Big\rangle_{N,t} =0.}

 The insertion of the operator  $\tr M^n =\sum_i \lambda_i^n$ can be 
realized by taking 
 a partial derivative with respect to the coupling $t_n$.
The  collective field \colf\ is therefore represented 
by the free  field operator acting on the coupling constants
\eqn\bsnf{
 {\bf \Phi}(z)= {1\over 2} \sum_{n=0}^{\infty} t_nz^n-    \ln 
 \Big({1\over z}\Big) {\p\over\p t_0}- 
\sum_{n=1}^{\infty} {z^{-n}\over n} {\p\over \p t_n}}
and the  loop equation \eqct\ is equivalent to the linear   condition
 \eqn\lolo{\oint_{\CC}
{dz'\over 2\pi i} {1\over z-z'}{\bf T}(z') \cdot  Z_N[t]
\equiv \big( {\bf T}(z)\cdot  Z_N[t] \big)_{<}=0}
imposed on the partition function, where 
\eqn\tsei{{\bf T}(z) = {1\over 4}:(\p {\bf \Phi}(z))^2:}
denotes the  energy-momentum  tensor for the gaussian field \bsnf . 
  Expanding 
  \eqn\currl{
  :\big(\p {\bf  \Phi}(z')\big)^2:=\sum_{n\in\Z} {\bf L}_n z^{-n-2}}
  we  find     a set of 
differential equations (Virasoro constraints)
 \eqn\viraa{ {\bf L}_n\cdot Z_N[t]=0 \ \ \ \ \ \  ( n\ge -1)}
 where
\eqn\lmOm{ 
{\bf L}_n=\sum_{k=0}^{n}   {\partial\over \partial t_k}
 {\partial\over \partial t_{n-k}}+
\sum_{k=0 }^{\infty} k t_k  {\partial\over \partial t_{n+k}}   }
 satisfy the algebra
  \eqn\viR{[{\bf L}_m,{\bf L}_n]=(m-n){\bf L}_{m+n}.}
 With respect to  the trivial variable $t_0$, we have the additional
 equation
 \eqn\trto{{\p \over \p t_0} Z_N [t]= N Z_N [t].}
 
 An elaborate analysis of the loop equations 
 and explicit expressions of the lowest orders of the $1/N$ expansion of 
the free energy are given in ref.
 \ref\ACKM{J. Ambjorn, L. Chekhov, C. Kristjansen and Yu. Makeenko, \np B 
 404 (1993) 127}.

   \subsec{ Orthogonal polynomials and  Hirota equations}
    
    It is   known \KMMM\   that the partition function  \hemm \  
 is a $\tau$-function of the KP 
hierarchy of soliton equations. 
 The global form of this hierarchy is given by 
the Hirota's   bilinear equations \hirota \ (for a review on the 
theory of the $\tau$-functions see, for example, \jimiwa.)
Below we will  derive the  Hirota equations  using 
 the formalism of    orthogonal 
polynomials.

 Let us   
 define  for each  $N\ \   (N=0,1,2,...)$ the polynomial
\eqn\polN{\eqalign{
P_{N,t}(\l)&= \langle \det (\l -M)\rangle  _{N, t} \cr
&=   \Big\langle  
\prod _{i=1}^N (\l-\lambda_i) \Big\rangle  _{N, t}\cr
&=\l^N -\langle \tr   M)\rangle  _{ N,t}\ \l^{N-1}+...+(-)^N
\langle \det  M)\rangle  _{ N,t}. \cr 
 }}
  It easy to prove that the polynomials \polN\  are orthogonal with 
respect to the measure $d\mu_t(\lambda)$.  
   Indeed,   
   \eqn\otT{\eqalign{
  & Z_N \int d\mu_t(\lambda_{N+1}) P_ {N,t}(\lambda_{N+1})\lambda_{N+1}^k  
\cr
 & \int \prod_{i=1}^{N+1}d\mu(\l_i) \Delta_{N+1}(\l_1,...,\l_{N+1})
  \Delta_N(\l_1,...,\l_N)\l_{N+1}^k\cr  
  &={1\over N+1}\int \prod _{i=1}^{N+1}d\mu_t(\l_i) 
\Delta_{N+1}(\l_1,...,\l_{N+1})\cr
&\sum_{s=1}^{N+1} (-)^{N+1-s}\l^k_s \Delta_{N+1}(\lambda_1,...,\hat 
\l_s,...,\l_{N+1}).\cr}}
 The sum in the integrand is the expansion of the determinant
 $$\left |
 \matrix{
 1 &\l_1& \cdots &\l_1^{N-1}&\l_1^k\cr
 1 &\l_2& \cdots &\l_2^{N-1}&\l_2^k\cr
  \cdots &\cdots     & \cdots  &\cdots &\cdots     \cr
 1 &\l_{N+1}& \cdots &\l_{N+1}^{N-1}& \l_{N+1}^k\cr
 }
 \right | $$
 with respect to its last column. It vanishes for $k=1,..., N-1$, which 
 proves the statement. For $n=N$, one finds
 \eqn\normp{Z_N\int d\mu_t(\l) P_{ N,t}(\l) \l^N= Z_{N+1}/(N+1).}
 Hence
  \eqn\Ort{
\int_{\lambda_L}^{\lambda_R}  
d\mu_t(\lambda)  \   
P_{ N,t}(\lambda)   P_{k,t}(\lambda)=\delta_{N,k} 
\ {Z_{N+1}[t] 
\over (N+1) Z_N[t]} . 
  }

The orthogonality   relations \Ort\ can be  written in  the form of a  
contour integral, 
namely, 
 \eqn\dDrd{
{1\over 2\pi i}\oint _{\cal C} dz\  \Bigg\langle \det 
(z-M)\Bigg\rangle_{ N,t}
\Bigg\langle {1\over \det (z-M)}\Bigg\rangle_{[k+1, t]}=\delta_{N,k}
   }
    where the integration contour $\CC$ encloses the point $z=0$ and the 
interval  
$[\lambda_L,\lambda_R]$.
Indeed,  the residue of  each of
 the $n$ poles is 
equal 
 to the left hand side of  \Ort \  multiplied by 
 $  Z_{k}[t] /  Z_{k+1}[t] $.

A set of more powerful  identities  follow from the fact that   the 
polynomial  $P_{N,t}(z)$
 is orthogonal to $any$ polynomial of degree less than $N$ and in 
particular to the  polynomials $P_{k,t'}[\l], k=1,2,...,N-1$
where $t'= \{ t'_n, n=1,2,...\}$ is another set   of coupling constants. 
Written in the form of contour integrals, these orthogonality relations 
state, for  $N'\le N$,
  \eqn\hirotd{
 \oint _{\cal C} dz\ e^{ \sum_{n=1}^{\infty}
(t_n-t'_n)z^n}
 \Bigg\langle \det (z-M)\Bigg\rangle_{ N,t}
\Bigg\langle {1\over \det (z-M)}\Bigg\rangle_{N', t'}=0 . 
    }
 
The Hirota equations  for the KP hierarchy are obtained from  eq. 
 \hirotd \  after expressing the mean values of $\langle \det^{\pm 
1}\rangle $ in terms of     the vertex 
operators
 \eqn\vop{
{\bf V}_{\pm}(z)= \exp\Big(  \pm \sum_{n=0}^\infty t_n z^n \Big) 
\exp\Big( \mp \ln {1\over z} \ {\partial\over \partial t_0}
\mp 
\sum_{n=1}^\infty 
{ z^{-n}\over n} 
 {\partial\over \partial t_n} \Big).
   }
It follows  from the     definition   \partf \ 
that  
\eqn\vtop{
{\bf V}_\pm(z)\cdot Z_N [t]
 =    e^{\pm \sum_{n=0}^\infty t_n z^n} \Big\langle  \det (z-M)^{\pm 
1}\Big\rangle_{ N,t} Z_N[t]
  }
 and eq.  \hirotd \ is therefore equivalent  to
\eqn\hireqn{
\oint_{\cal C}  dz \   \Big({\bf V}_+(z)\cdot Z_N[t] 
\Big)\ 
\Big({\bf V}_-(z) \cdot Z_{N'}[t']\Big) 
=0  \ \ \ \ \ \ (N'\le N ),
   }
    which is one of the forms of the Hirota equation for KP \jimiwa .
   
    After a change of variables  
\eqn\hireqys{
x_n= {t_n+t'_n\over 2}, \ \ y_n= {t_n-t'_n\over 2},
 } 
the Hirota equations  \hireqn \ take its  canonical form  
\eqn\hirver{
  {\rm Res}_{z=0}  \ z^{N-N'}
 e^{   \sum _{n=1}^{\infty}  2y_n z^n }e^{-\sum 
_{n=1}^{\infty}
  {z^{-n}\over n} {\partial\over \partial y_n} } Z_N[x+y]Z_{N'}[x-y]=0}
  where $ N'\le N$.
The differential equations of the KP hierarchy are obtained by expanding 
 \hirver \ in $y_n$.    
For example, for $N'=N$,  the   
coefficient in front of $y_1^4$  is 
\eqn\olki{
\Big({\partial^4\over\partial y_1^4}+3{\partial^2\over\partial y_2^2}-4
{\partial\over\partial y_1}{\partial\over\partial y_3}\Big)Z_N[t+y]
 Z_{N'}[t-y]\Big|_{y=0}=0
  }
and one finds  for the ``free energy'' $u[t]= 2 
{\partial^2\over\partial t_1^2}
\log Z_N$ the KP equation
\eqn\KPe{
3{\partial^2 u\over\partial t^2_2} + {\partial\over\partial t_1}
\Big[ -4 {\partial u\over\partial t_3} +6 u {\partial u\over\partial t_1}
+{\partial^3 u\over\partial t_1^3}\Big]=0.
  }
 The lowest equation in the case   $N'=N-1$ (modified KP) 
 is the so called Miura transformation 
relating the 
functions $u=2\partial_1^2 \log Z_N[t]$ and $v= \log (Z_{N+1}[t]/Z_N[t])$: 
  \eqn\Me{u=
 \partial_2 v-\partial_1^2 v -(\partial_1v)^2 .
  }
  One can check directly that these differential equations are satisfied 
by the
    asymptotic  expansion around the gaussian point 
 \eqn\asser{
 Z_N[t]= (-t_2)^{-N^2/2} \exp \Big( Nt_0 -{N\over 4} {t_1^2\over t_2}+
{N^2\over 4} {t_3 t_1\over t_2^2} - {N \over 
 8} {t_1^3 t_3\over t_2^3}+...\Big).
 }

  The Hirota equations themselves are not sufficient to determine 
  uniquely the partition functions $Z_{N}[t]$.   The additional 
  information comes from the ``string equations''
  \eqn\streqa{\sum_n nt_n\int d\mu _t( \l) \l ^{n-1} P_{ N,t}(\l) 
   P_{N-1+\sigma,t}(\l) = \delta_{\sigma, 0}
  {Z_N\over Z_{N-1}} \ \ \ \ (\sigma = 0,1)}
 obtained,  with   integrating 
   by parts,  from 
\eqn\steq{ \int d\mu_t(\l) P_{N-1+\sigma ,t}(\l) \p_{\l}P_{ N,t}(\l)
 =\delta_{\sigma,0} {Z_N\over Z_{N-1}}\ \ \ \ (\sigma = 0,1)}
and therefore  valid  only if  $d\mu(\l) /d\l$ vanishes at the endpoints 
$\lambda_L, \l_R$ of the eigenvalue interval.       

   In terms of  vertex operators, the string   equations \streqa\  read
  \eqn\strea{\sum_n nt_n
\oint _{\cal C} dz\ z^{n-1}\ ( {\bf V}_-(z)\cdot 
Z_{N+\sigma}[t] ) ( {\bf V}_+(z)\cdot Z_N[t])  =\delta_{\sigma,0} N\ 
Z_N[t]^2
     }
     with $\sigma = 0,1$.

 \newsec{ The $O({\rm n})$ model}
The partition function of the  $O({ \rm n}) $ matrix model\foot{ We use a  
Roman 
letter for the parameter n$\in[-2,2]$  to avoid confusion with the  
summation index $n$ running the set of natural numbers.}    is defined by 
the  $N\times N$  matrix integral \onon
 \eqn\mort{
 Z_N ^{ O({\rm n})} [t ]\sim 
 \int dM  \exp\Big[ \sum_{n=0}^{\infty}
 t_n \ {\rm tr}M^n 
+ {{\rm n}\over 2}
 \sum_{n+m\ge 1 } {T^{-n-m}\over n+m} \
 {(n+m)!\over n!\ \ m!} {\rm tr}M^n 
 {\rm tr}M^m 
 \Big].
  }
   (the sum over $n$ and $m$ runs the set of nonnegative integers) and  
describes the  ensemble of nonintersecting loops on a 
random graph. A loop of length $n$  is weighted by a factor  ${\rm  {n}} 
T^{-n}$ where 
  $T$ is the   temperature of the loop gas.
 For  n$\in [-2,2]$, the model exhibits    critical behavior  with 
spectrum of the central charge   $C=1-6(g-1)^2/g$,  $g=  {1\over \pi } 
\arccos (-{\rm n}/2)$;   the critical behavior for $ |{\rm n}|>2$ 
is this of a branched polymer \berchar .
   The  matrix integral  \mort \ 
can be again interpreted as the partition function of a Coulomb gas with 
  more complicated but still pairwise  interaction between particles:
\eqn\pfon{
 Z_N ^{ O({\rm n})} [t ]= \ T^{-{1\over 2}{\rm n}N^2} \int   
\prod _{i=1}^N    d\lambda_i 
\ \exp 
\Big( \sum_{n=0}^{\infty} t_n \lambda^n_i\Big)  \  {\prod_{1\leq i<j \leq 
N} (\lambda_i - \lambda_j)^2 \over
 \prod_{i,j} (T-\lambda_i-\lambda_j )^{{\bf\rm n}/2}}. 
   }
 The   interval of integration
$[\lambda_L, \lambda_R]$   should be such that   $\lambda_R\le T/2 $.  
With 
the last restriction the  denominator in the integrand never vanishes.
 The choice of the integration interval does not influence the 
quasiclassical
 expansion and hence the geometrical interpretation in terms of a  gas 
 of loops on the random planar graph.  
 The saddle point spectral density  is  automatically supported by an
  interval on the half-line
  $[-\infty, T/2]$ \onon.  The most natural choice for the eigenvalue 
interval 
  is therefore $\l_L\to-\infty, \l_R\to T/2$, with the conditions 
  $d\mu(\l)/d\l|_{-\infty}=d\mu(\l)/d\l|_{T/2}=0$.

The   $O({\rm n} )$ model reduces to the hermitian matrix model in the 
limit n$\to 0$ and/or $T\to\infty$, and 
can be considered as a deformation of the latter in the 
following sense.
Let us define  the differential operator
\eqn\perth{
{\bf H}=   { {\rm n}\over 2} \Bigg[ -\ln T {\partial^2\over\partial t_0^2} 
+\sum_{n+m\ge 1 } {T^{-n-m}\over n+m} \
  {(n+m)!\over n!\ \ m!} {\partial\over\partial t_n}
{\partial\over\partial t_m} \Bigg] 
}
acting on the coupling constants. It is easy to see that 
  the  partition function \pfon\ is obtained from the 
partition function of the one-matrix model by acting with the operator
 $e^{\bf H}$:
  \eqn\oppq{
   Z_N ^{ O({\rm n})}  [t]=  e^ {\bf H} \cdot Z_N[t] .
}
This simple observation will be of crucial importance for our further  
consideration. It means that  the integrable 
structure  of the 
one-matrix model survives in some form in the 
 $O(n)$ model.   
 The operator \perth\ defines a  canonical  transformation
 \eqn\cantr{\eqalign{
 {\p\over\p t_n}& \to {\p\over\p t_n} ,\cr  
 t_n &\to \tilde t_n = t_n +{\rm n} \sum _{ m  }  T^{-n-m} 
  {(n+m-1)!\over n!\ \ m!} {\p\over\p t_m}\cr}}
   preserving the symplectic structure.  

 \subsec{Loop equations and Virasoro constraints} 

 From the translational invariance of the integration measure $d\lambda_i$ 
we find
\eqn\dsaon{
\Big\langle \Big(\sum_{i=1}^N{1\over z-\lambda_i} \Big)^2+  {\rm n}
\sum_{i,j}  { 1\over z- \lambda_i }
{ 1\over T-\lambda_i-\lambda_j}+ \sum_{n\ge 0}{1\over 
z-\lambda_i}nt_n\lambda_i^{n-1}  
\Big\rangle_{N,t} =0.}
or, expressing  the sum as  a contour integral,  
\eqn\eqnct{\Big\langle W(z)^2 + \oint_{\CC_-}
{dz'\over 2\pi i} {1\over z-z'}W(z')[ {\rm n}W(T-z' )+\sum_n
nt_n (z')^{n-1}]\Big\rangle_{N,t} =0.}
The integration contour $\CC_-$ encloses the interval $[\l_L,\l_R]$ and
 leaves outside the interval
$[T-\lambda_R,T-\lambda_L]$ and the point $z$.
 Introducing the collective field
 \eqn\colfo{\tilde \Phi(z)=\tr \log (z-M) -{{\rm n}\over 2}\tr \log (T-z-M)
 +{1\over 2}  \sum_n t_n z^ n }
we write the last equation as 
\eqn\eqct{  \oint_{\CC_-}
{dz'\over 2\pi i} {1\over z-z'} \Big\langle \big(\p \tilde 
\Phi(z')\big)^2  
\Big\rangle_{N,t} =0.}
The    field \colfo\ is   represented 
by the linear   operator 
\eqn\bsnfon{   
 \tilde {\bf \Phi}(z)
 ={1\over 2} \sum_{n=0}^{\infty} t_nz^n-    \ln 
 \Big({ (T-z)^{{{\rm n}\over 2}}\over z}\Big) {\p\over\p t_0}- 
\sum_{n=1}^{\infty}  {z^{-n}-{{\rm n}\over 2}(T-z)^{-n}\over n }
 {\p\over \p t_n} }
 and is related to the ``bare'' field \bsnf \  by 
  $\tilde {\bf \Phi}(z)=  e^{{\bf H}}{\bf \Phi}(z)e^{-{\bf H}} $.
      The loop equation \eqct   \ is therefore  equivalent to the 
  linear differential constraints
  \eqn\vironn{\tilde {\bf L}_n\cdot Z^{O({\rm n})}[t]=0 \ \ \ \ \ (n\ge 
-1)}
  where the operators 
 \eqn\yoon{\eqalign{
\tilde {\bf L}_n=&\sum_{k=0}^{n}   {\partial\over \partial t_k}
 {\partial\over \partial t_{n-k}}+
\sum_{k }k t_k  {\partial\over \partial t_{n+k}} \cr & + 
 {\rm n}  \sum_{ k,m} T^{-k-m-1} 
 { (k+m)!\over k!\ m! }  {\partial\over \partial t_{n+k+1}}
 {\partial\over \partial t_{m} } 
 \cr}}
  are related to the standard   Virasoro generators \lmOm \ by
  $ \tilde {\bf L}_n = e^{{\bf H}} {\bf L}_ne^{-{\bf H}}$ and therefore 
  satisfy the same algebra.

 \subsec{ Bilinear functional equations}

    In quite similar way     
   the    Hirota  equations   \hireqn\ provide, due to  the relation \oppq 
,
   a set of  bilinear equations   for the 
partition function of the 
$O({ \rm  n})$ model
 \eqn\hireqo{{
 \oint _{\cal C_-} dz   \Big( 
  \tilde {\bf V}_+ (z)\cdot   Z _N^{ O({\rm n})} [t] 
\Big)\ 
\Big( \tilde  {\bf V}_-  (z)\cdot    Z _{N'}^{ O({\rm 
n})}[t']\Big) 
=0  \ \ \  \ \ \  
(N'\le N) 
}}
 where $ \tilde {\bf V}_{{\pm} }   (z)$
are the transformed   vertex operators
 \eqn\tvop{\eqalign{\tilde {\bf V}_{{\pm} } &=
  e^{ {\bf H}} {\bf V} _{\pm}(z) e^{-\bf H}\cr
  &= 
( T- 2z)^{-{\rm n}/2} 
 \exp \Bigg( \pm \sum_{n=1}^{\infty} t_n z^n \Bigg)\cr
\exp \Bigg( &\mp \ln \Big[{ (T-z)^{\rm n}
\over z} \Big] {\partial\over\partial t_0}  \mp  \sum_{n=1}^{\infty} 
{z^{-n}- { \rm n}  (T-z )^{- n} 
\over n} \ {\partial \over \partial t_n}   \Bigg).
  \cr} 
 }
  After being 
expanded in $t_n-t_n' $, eq.  \hireqo \ generates   a hierarchy 
   of differential equations, each of them  involving derivatives with 
respect to an infinite
    number of ``times" $t_n$.

The functional relations \hireqo \
  are equivalent to the   bilinear  relations 
\eqn\hiron{\eqalign{
 & \oint_{\cal C_-} {dz\over (T-2z)^{\rm n}} \exp\Big(\sum 
_{n=1}^{\infty}
  	(t_n-t'_n) z^n \Big) \cr
&  	\Bigg\langle   {\det(z-M) \over
\det(T-z-M)^{\rm n}} \Bigg\rangle_{ N,t}  \Bigg\langle
 {  \det (T-z-M)^{\rm n}\over
\det(z-M)}\Bigg\rangle_{[N',t']}=0 \ \ \   \ \ \ \ \ (N'\le N)\cr}
 }
  where   $\langle \ \ \rangle_{ N,t}$ denotes  the average  
corresponding to the partition function  \pfon . 
 These relations can be also
proved directly by  exchanging the order of the integration in the $\l$'s 
and the  contour integration in $z$.
The    "string equations", needed to determine completely the partition 
function, are obtained from   \strea \
  by   replacing 
$ {\bf V}_{\pm}  \to  \tilde {\bf V}_{\pm} (z)$.

       \newsec{  $ADE$ and $\hat A\hat D\hat E$ models}

    The   $ADE$ and  $\hat A\hat D\hat E$  matrix
      models   give a nonperturbative microscopic realization of the 
      rational string theories with $C\le 1$.
     Each one of these models is associated with a rank $r$    classical 
simply laced  
     Lie algebra (that is, of type  $A_r, D_r, E_{6,7,8}$) or its afine 
extension, and represents a system of $r$ coupled random matrices.  The 
matrices
      $M_a$ of size   $N_a\times N_a \ (a=1,2,...,r)$ are associated 
with     the nodes of the Dynkin  diagram of the simply laced Lie algebra, 
the    latter being  defined by   its adjacency matrix $G$ with elements 
  \eqn\adJ{
G^{ab}=\cases{1 & if the two nodes are the extremities of a link    $<ab>$ 
\cr
0& otherwise.\cr}
 } 

The partition function $Z^{  G}_{\vec N}    [ \vec t]$ depends on $r$ sets 
of coupling constants $\vec t = \{t_{n}^a\ |\ a=1,...,r;\ 
n=1,2,...\}$. 
The interaction is of
nearest-neighbor type and the   measure is a product of factors associated 
with the nodes $a$ and the    links $<ab> $ of the Dynkin 
diagram
\eqn\DdD{\eqalign{ 
Z^{  G}_{\vec N}    [ \vec t] \sim  & \int \prod_{a=1}^r dM_a \exp 
\Bigg(
 \sum_{a=1}^r \sum_{n=1}^{\infty}  t_{n}^a \ \tr M_a^n
\cr & + {1\over 2} \sum_{a,b} G^{ab}   \sum_{m,n=1}^{\infty} 
{T^{-m-n}\over m+n} 
{(m+n)!\over 
m!\ n!}\ \tr M_a^m \ \tr M_b^n \Bigg). \cr}
 }
 The target space of the $A_r$ model is an open chain of $r$ points and 
 the one of  the $\hat A_{r-1} $ model   is a circle with $r+1$ points.  
  In this sense the  $O(2)$ model can be referred to as the  $\hat A_0$
 model of  the $\hat A$ series. The continuum limit of the $\hat A _r$ 
model
 is that of a free field coupled to gravity and compactified at the radius 
$r$
in a  scale where the  self-dual radius is $r=2$.  
 
 Again, the   only nontrivial integration is with respect to the  
eigenvalues 
$\lambda_{ai}  \ ( i=1,...,N_a)$ of the 
matrices $M_a$:
 \eqn\mtriaa{ 
Z^{G}_{\vec N}[\vec t] = \prod _{ a=1}^r\prod_{i=1}^{N_a} d \lambda_{ai}  
 e^{\sum_n t_{n}^a \lambda_{ai }^n } { { \prod_a \prod_{i< j} 
 (\lambda_{ai }-\lambda_{aj} )^2 \over \prod_{<a b>}
   \prod_{i,j}  (T- \lambda_ {ai} -\lambda_ {bj} )}}.
  }  
  The  domain of  integration  is assumed to be   a  
compact  interval $[\lambda_L, \lambda_R]$ with $\lambda_R\le T/2$.

The partition function 
 \mtriaa \ 
can be obtained by acting on 
 the product of $r$ one-matrix partition functions  $  Z_{N_a}[t^a], 
a=1,...,r,$
with the  exponent of the second-order differential operator  
\eqn\jJj{
 {\bf H} = {1\over 2} \sum_{a,b}G^{ab}
   \Bigg[   \ln T^{-1}  {\partial\over \partial t^a_{0}}
 {\partial\over \partial  t^b_{0}} +
\sum_{n+m\ge 1 } {T^{-n-m}\over n+m} \
 {(n+m)!\over n!\ \ m!}  {\partial\over \partial t_{ n}^a}
 {\partial\over\partial t^{ b}_m}\Bigg] ,  
 } 
 namely,
\eqn\xprd{
  Z^G_{\vec N}    [ \vec t]= e^{{\bf H}} \cdot \prod_{a=1}^r 
Z_{N_a}[t^a].
       }
       
        \subsec{Loop equations and Virasoro constraints}

 The Dyson-Schwinger  equations are written in terms of the  
$r$-component  collective  variable
 \eqn\ohh{
W_a(z)=     \sum_{i=1}^{N_a} {1\over 
z-\lambda_{ai}}. }
 They are obtained as in the previous cases, by  shifting the variable 
$\lambda_{ai}$, and read
\eqn\ADEeqa{\Big\langle W_a(z)^2 + \oint_{\CC_-}
{dz'\over 2\pi i} {1\over z-z'}W(z')\Big[ \sum_bG^{ab}W_b(T-z' )+\sum_n
nt_n^a (z')^{n-1}\Big]\Big\rangle_{N,t} =0.}
The loop equations  are equivalent to the differential constraints 
\eqn\virara{ \tilde {\bf L} ^a_n\cdot  Z^G[t] =0 \ \ \ \ \ \ ( n\ge -1, \  
a=1,...,r),}
where the operators
\eqn\yayI{\eqalign{\tilde 
{\bf L}_n^a=&\sum_{k=0}^{n}   {\partial\over \partial t_k^a}
 {\partial\over \partial t_{n-k}^a}+
\sum_{k }k t_k^a  {\partial\over \partial t_{n+k}^a} \cr & + 
\sum_b 
G^{ab} \sum_{ k,m} T^{-k-m-1} 
 { (k+m)!\over k!\ m! }  {\partial\over \partial t_{n+k+1}^a}
 {\partial\over \partial t_{m} ^b} 
 \cr}} 
    are related to the ``bare''  Virasoro generators   by $\tilde {\bf 
L}_n^a=
e^{{\bf H}} {\bf L }_n^{a}(z)  
  e^{{-\bf H}}$
and  form $r$    commuting  Virasoro  algebras
    \eqn\vitcm{[\tilde {\bf L}_m^a, \tilde {\bf 
L}_n^b]=\delta_{a,b}(m-n)\tilde {\bf L}_{m+n}^a.}
 With respect to the trivial variables $\vec t_0$, we have the additional equations
 \eqn\adtr{{\p\over\p t^a_0} Z^G[t]=N_a Z^G[t], \ \ \ \  (a=1,..., r).}
       
       \subsec{Bilinear functional equations}
  Each of the one-matrix partition functions on the  right hand side of 
  \xprd \
  satisfies the Hirota equations   \hireqn . As a consequence, the  left 
hand 
  side satisfies  a set  of $r$  
functional equations associated with the nodes $a=1,...,r $ of the Dynkin 
diagram.
Introducing the transformed vertex operators
 \eqn\gmma{
    \tilde {\bf V }_{\pm}^{a}(z) 
 =e^{{\bf H}} {\bf V }_{\pm}^{a}(z)  
  e^{{-\bf H}} 
    }
 we find, for each node  $a=1,...,r,$
 \eqn\hireqoa{
 \oint _{\cal C_-} dz   \Big( 
\tilde {\bf V }_+^{a}(z)\cdot  Z _{\vec N}^{G}[\vec t] \Big)\ 
\Big( \tilde {\bf V }_-^{a}(z)\cdot  Z _{\vec N'}^{G}[\vec t']  \Big) =0 \ 
\ \  \ \ \ 
(N'_a\le N_a)
 }
 where the 
  integration contour
${\cal C_-}$ in 
 \hireqoa \ encloses the interval $[\lambda_L, \lambda_R]$ but leaves 
outside the interval $[T-\lambda_R, T-\lambda_L]$.

 Inserting \jJj\  in the definition  \gmma\ we 
 find the explicit form
  of the  dressed vertex  operators    
\eqn\mvoA{\eqalign{
  & \tilde {\bf V }_{\pm}^a(z)  = \cr
& \prod_{b} (T-2z)^{-{1\over 2}G^{ab}} e^{ \pm \sum_{n=0}^{\infty} 
t^a_n z^n}
 \exp \Bigg(\mp\Bigg[ \ln z^{-1}  {\partial \over 
\partial t^a_0}
+  \sum_{n=1}^{\infty} {z^{-n}\over n}{\partial \over 
\partial t^a_n}\Bigg]\Bigg) \cr 
& \exp \Bigg(\pm \sum _b G^{ab} \Bigg[  \ln (T-z)^{-1} {\partial \over 
\partial t^b_0}
   +  \sum_{n=1}^{\infty} { (T-z )^{- n} \over n} 
{\partial \over \partial t^b_n} \Bigg]
 \Bigg) \cr}
}
  and write   the   generalized Hirota equations \hireqoa \     in terms 
of  correlation functions of 
determinants:  
     \eqn\hironA{\eqalign{ 
 &\oint_{\cal C_-} dz  
{e^{\sum_{n=1}^{\infty} [t^a_n -t'^a_n ]z^n  }\over
 \prod_{b} (T-2z)^{  G^{ab} } }
  	\Bigg\langle   {\det (z-M_a)\over
\prod_b  \det (T-z-M_b)  ^{G^{ab}}} \Bigg\rangle_{\vec N,\vec t}  \cr
& \Bigg\langle
{   \prod_b   \det(T-z-M_b)^{ G^{ab}}\over\det (
z-M_a)}\Bigg\rangle_{[\vec N',\vec t']}=0\ \ \ \ \ \ \ \ \ \ \  \ \ \ 
(N_a'\le N_a) . \cr}}
  Here $\langle \ \ \rangle_{\vec N,\vec V}$ denotes  the average in the 
ensemble 
  described by the partition function  \mtriaa  .  
  It is possible to prove eq.  \hironA \   directly by 
  performing the contour integration. It is essential for the proof  that 
 the interval $[T-\lambda_R, T-\lambda_L]$ is outside  the contour 
$\cal C_-$.
  
\newsec{Continuum limit}

  We  conclude with a  brief description of  the continuum limit of the 
above
construction.
In the scaling limit the field \bsnf\ is expanded in the half-integer 
powers 
of the (shifted   and rescaled) variable $z$
 \eqn\bsnfc{
\p{\bf \Phi}(z)= {1\over 2} \sum_{s>0  }   t_sz^{ s } 
-\sum_{s<0 }  {z^{-s  }\over  s} {\p\over \p t_s}\ \ \ \ \  (s\in 
\Z+{1\over 2})}
where  $t_{1/2}, t_{3/2},...$ are the corresponding coupling constants. We 
adopted half-integer indices 
in order to follow more easily the analogy with the microscopic formulas.
  The loop equations \lolo\  lead to the 
Virasoro constraints 
 \eqn\viraa{ {\bf L}_n\cdot Z_N[t]=0 \ \ \ \ \ \  ( n\ge -1)}
 with
\eqn\lmOm{ 
{\bf L}_n=\sum_{s+s'=n}   {\p\over \p t_s}
 {\p\over \p t_{s'}}+
\sum_{s-s'=n}  s t_s  {\p \over \p t_{s'}} +
{1\over 4} \sum_{-s-s'=n} ss't_st_{s'} . }
Note that in the continuum limit the string equation 
  is  equivalent to the constraint 
 defined by the operator ${\bf L}_{-1}$ and the constraints \viraa\
 determine completely the partition function.
  The Hirota equations for the KdV hierarchy 
are formulated in terms of  the vertex operators 
 \eqn\vopc{
{\bf V}_{\pm}(z)= \exp\Big(  \pm \sum_{s\in \Z_++{1\over 2} } t_s z^s 
\Big) 
\exp\Big(  \mp 
\sum_{s\in \Z_--{1\over 2}} 
{ z^{-s}\over s} 
 {\partial\over \partial t_s} \Big).
   }

 The  loop equations and the  bilinear functional  equations 
in the continuum limit of the $O({\rm n}) $ model are obtained 
by transforming the Virasoro generators and the vertex operators with the 
operator $e^{{\rm H}}$, where ${\rm H}$ is given by  
 \eqn\jJoo{
 {\bf H} = {{\rm n} \over 2}  \sum_{s+s'\ge 1 } {M^{-s-s'}\over s+s'} \
 {\Gamma (s+s'+1)\over \Gamma(s+1)\ \Gamma (s'+1)}  {\partial\over 
\partial t_{ s}}
 {\partial\over\partial t_{s'}}. 
 } 
 The parameter $M$ measures the deviation from the critical point  
 (we use the same notation as in \Iade) so that  $1/M$ is  the inverse  
correlation length of the matter field. In the limit $M\to \infty$ the 
deformation  operator $e^{{\rm H}}$ becomes the identity operator.

Similarly, the statistical model associated with the adjacency matrix 
$G_{ab}$ is described by 
the deformation operator  
 \eqn\jJjc{
 {\bf H} = {1\over 2} \sum_{a,b}G^{ab}
   \Bigg[   
\sum_{s+s'\ge 1 } {M^{-s-s'}\over s+s'} \
 {\Gamma (s+s'+1)\over \Gamma(s+1)\ \Gamma (s'+1)}  {\partial\over 
\partial t_{ s}^a}
 {\partial\over\partial t^{ b}_{s'}}\Bigg] .
 } 
acting on the set of coupling constants
\eqn\ccc{t=t^a_s, \ \ \ \ (a=1,...,r; \ s\in \Z_++{1\over 2}) .}
 
 The partition function   is
 determined by the Hirota equations and $r$ the subsidiary conditions
 (string equations)
 \eqn\strqq{ \tilde {\bf  L}^a_{-1}\cdot   Z^G[\vec t]=0 \ \ \ \ 
 \ \ \  (a=1,..., r).}

\newsec{Concluding remarks}

Our construction represents an alternative   nonperturbative formulation  
of 2D gravity 
suited for investigating the infrared properties of the theory.
At very large distances the fluctuations of the matter fields can be 
neglected 
and  the theory decouples to $r$ copies of pure gravity, where $r$ is the 
number of points of the target space. At smaller distances the 
fluctuations become important, 
but still can be considered as a perturbation. A highly nontrivial fact, 
related to the
specific realization of 2D gravity we are considering, is that the 
perturbation can be imposed by means of a differential operator acting on 
the coupling constants.  As a consequence of this, the  Virasoro  
constraints and the  Hirota bilinear equations characterizing   pure 
gravity  survive in some form in  the  theory with matter fields.

The  coupling constants \ccc\ are related to the coupling constants 
associated with scaling operators as follows.
The coupling constants $t^a_s$ are the coefficients in the expansion of 
the loop fields
 (always in   half-integer powers of $z$) at $z=0$, which is   convergent 
in the circle $|z|<2M$.   The  coupling constants associated with scaling 
operators  represent the coefficients in the expansion (in fractional 
powers of $z$) of the same loop fields at $z=\infty$, which is   
convergent  in the domain $|z+M|>M$.
 
 Most likely   the two realizations of the   $A_r$ 
series, the one which we are considering and  and  the one   as a matrix 
chain with interaction $\tr M_aM_{a+1}$,
have the same scaling limit. We therefore expect that both cases are  
described, 
in the continuum limit, by $(r+1)$-reduced  KP hierarchy and  the  
differential constraints  generating a  $W_{r+1}$ algebra.  
      
  The exact integrable structures associated with the  $\hat A_r $ models 
are discussed in \Itrieste. It is shown there that the canonical  
partition function of the $\hat A_0$ model
is a $\tau$-function   of the KdV hierarchy. 
Thus the KdV hierarchy appears in two cases: in pure 
gravity (the $A_1$ model) and in the case of a gaussian field compactified 
at the Kosterlitz-Thouless radius (the  $O(2)\equiv \hat A_0 $ model).
One can speculate that  the  partition functions of the  $A_r $ and $\hat 
A_{r-1}$ models are  $\tau$-functions of the  $(r+1)$-reduced    KP 
hierarchy and that, more generally,
  the integrable structures associated with  the $ADE$ and $\hat A \hat D 
\hat E$ 
models are given by the exceptional 
  hierarchies studied in \KW .

  \listrefs
  \bye